\documentclass[reprint, amsmath,amssymb, aps,pra]{revtex4-1}

\usepackage[utf8]{inputenc}
\usepackage[draft,inline,nomargin]{fixme}
\usepackage{graphicx}
\usepackage{dcolumn}
\usepackage{bm}
\usepackage{braket}
\usepackage{physics}

\begin{document}

%\preprint{APS/123-QED}

\title{Cooling in a parametrically driven optomechanical cavity}%

\author{Pablo Yanes-Thomas}
\affiliation{Instituto de Investigaciones en Matem\'aticas Aplicadas y en Sistemas, Universidad Nacional Aut\'onoma de M\'exico,
C.P. 04510, Ciudad de M\'exico, M\'exico}%

\author{Marc Bienert}
\affiliation{Sch{\"u}lerforschungszentrum S{\"u}dw{\"u}rttemberg, D-88348 Bad
  Saulgau, Germany}

\author{Pablo Barberis-Blostein}

\affiliation{Instituto de Investigaciones en Matem\'aticas Aplicadas y en Sistemas, Universidad Nacional Aut\'onoma de M\'exico,
C.P. 04510, Ciudad de M\'exico, M\'exico}%

\date{\today}

\begin{abstract}

  We obtain a master equation for a parametrically driven
  optomechanical cavity. We use a more correct dissipation model that
  accounts for the modification of the quasi-energy spectrum caused by
  the driving. When the natural frequency of the mechanical object
  oscillates periodically around its mean value, the master equation
  with the improved dissipation model is expressed using Floquet
  operators. We apply the corresponding master equation to model the
  laser cooling of the mechanical object. Using an adiabatic
  approximation, an analytical expression for the number of
  excitations of the mechanical oscillator can be obtained. We find
  that the number of excitations can be lower than in the non-time
  dependent case. Our results raise the possibility of achieving lower
  temperatures for the mechanical object if its natural frequency can
  be controlled as a function of time.
\end{abstract}

\pacs{Valid PACS appear here}% PACS, the Physics and Astronomy
\maketitle

%\tableofcontents

\section{Introduction}

Quantum cavity optomechanics studies systems composed of macroscopic
mechanical objects, such as mirrors, and an optical cavity's quantized
light field, coupled via radiation pressure. In a common
scheme one of the end-mirrors of a Fabry-Perot cavity is suspended
while able to freely oscillate. When photons are reflected by the
mirror, there is a momentum transfer between the light field and the
mirror; as the cavity's resonance depends on its length, the
mechanical displacement in turn affects the light field inside the
cavity. Some of the first theoretical work predicting this sort of
coupling, between light and mechanical object, is described in
\cite{BraginskiiOG}. This interaction between the macroscopic
mechanical object and the light field leads to several interesting
effects such as optomechanically induced transparency \cite{WeissOIT},
the optical spring effect \cite{VogelOT} or, most relevant to this
study, optomechanical cooling\cite{CohadonCM, CorbittOC, SchliesserRPC, LCNooshi}, which was first proposed by Mancini, et
al \cite{ ManciniOC}.
	
Optomechanical cooling consists of the damping of the end mirror's
mechanical motion due to the radiative coupling to the cavity field.
Sideband cooling takes place when the cavity's resonance is much
narrower than the mechanical frequency. It can be understood as Raman
scattering of incident photons \cite{MarquardtQTOQ} which are
red-detuned from the cavity resonance. When the parameters are chosen
appropriately, incident photons absorb a phonon from the
mechanical oscillator in order to scatter into the cavity's resonance
mode, resulting in cooling of the resonator. For coherent quantum
control over a mechanical object, it must be close to a pure quantum
mechanical state \cite{KippenberCO}, so effective methods of cooling
macroscopic objects to low temperatures are highly desirable.

One possible avenue for manipulating the mechanical object and
improving cooling lies in controlling the mechanical resonator's
frequency as a function of time \cite{JockelMR}. There have been other
studies that include the modulation of optomechanical parameters. Some
of these include modulating the spring constant and the interaction
strength to achieve a splitting of the cavity
sidebands \cite{AranasSlowModulation2016} reach a non-linear quantum
regime  \cite{YinNonlinearEffectsModulated2017}, or achieve
controllable quantum squeezing \cite{ZhangModulationSqueezing2018}.
Other studies cover the periodic Langevin equations that arise in a
bi-chromatically driven optical
cavity \cite{MalzBiChromaticallyDriven2016} and modulating the
amplitude of the driving field \cite{MariGentleModulation2009} in order
to achieve squeezing of the mechanical resonator. The effect of a
modulated spring constant on the mechanical object's final temperature
was studied in \cite{BarberisLC}. In that study it was found that the
final temperature of the parametrically driven harmonic oscillator was
larger than the non-driven case. However, the master equation from
which the cooling rates were derived accounted for the natural
frequency of the mechanical resonator via ad-hoc time-dependent
coefficients that were introduced after performing the Markov
approximation. In this paper we extend the study of the cooling
dynamics to the case when the time-dependence of the system is taken
into account in the derivation of the master equation.

The formalism we apply (section \ref{OptmechH}) is based on Floquet
theory and was demonstrated to be a more accurate treatment
\cite{HanngiFM}. For the case where the drive consists of a small
periodic oscillation with respect to the central frequency of the
mechanical oscillator, the Floquet operators can be given explicitly
(section \ref{SolSmallOsc}). Under the adiabatic approximation, we
derive an approximated expression for the mean mechanical excitation
number in the final stages of optomechanical cooling and compare this
prediction to the time-independent case (section \ref{LasCool}). From
the treatment presented here, it follows that lower temperatures can
be obtained if the mechanical object is parametrically driven. Our
result suggests that the details of the theoretical dissipation model
for the mechanical oscillator can have a significant influence on the
resulting temperature (section \ref{ConCl}).

\section{Optomechanical Hamiltonian}\label{OptmechH}
\subsection{Hamiltonian with Floquet Operators}
	
The Hamiltonian for a parametrically driven optomechanical system
\cite{LCNooshi}, in a reference system that rotates with the same
frequency as a laser that continuously pumps photons into the cavity
is

\begin{equation}
H(t) =   H_{\rm cav} + H_{\rm mec}(t) + H_{\rm int} + H_{\rm pump},\label{eq:Hfull}
\end{equation}

where

\begin{align}
H_{\rm cav} =& -\hbar \delta a^\dagger a,\\
H_{\rm mec}(t) =& \frac{p^2}{2M} + \frac{1}{2}M \nu^2 (t) x^2,\label{eq:Hmec}\\
H_{\rm int} =& -\hbar g_c a^\dagger a x,\\
H_{\rm pump} =& \hbar\frac{\Omega}{2}(a^\dagger + a),
\end{align}
where $\delta = \omega_{laser} - \omega_{cav}$ is the detuning between
laser and cavity. The mechanical oscillator's mass is denoted by $M$,
$p$ and $x$ are its momentum and position operator, and $\nu(t)$ the
modulated mechanical frequency. The term $H_{\rm int}$ models the
interaction between the field and the cavity mirror where $g_c$ sets
the strength of the coupling \cite{KippenberCO}. Finally,
$H_{\rm pump}$ describes the pumping of the cavity by a field with
strength proportional to $\Omega$. The mechanical oscillator
Hamiltonian has an explicit time dependence given by $\nu(t)$. We
assume here a periodic function of time, which allows us to employ the
Floquet formalism.

The Floquet operators are analogous to the usual creation and
annihilation operators for the standard harmonic oscillator and can be
expressed in terms of the mechanical oscillator's position and
momentum operators \cite{HanngiFM}. These operators are

\begin{equation}\label{FloquetOperators}
\Gamma(t) = \frac{1}{2i}\left[\hat{x}\sqrt{\frac{2M}{\hbar}}\dot{f}(t)-\hat{p}\sqrt{\frac{2}{M \hbar }}f(t)\right],
\end{equation} as well as its Hermitian conjugate. $f(t)$ is the solution to the classical time-dependent harmonic oscillator equation of motion in one dimension 

\begin{equation} \label{TimeDependentHO}
\ddot{f} + \nu(t)^2f=0,
\end{equation} and is generally a complex function \cite{BrownPT}. This equation has two solutions \cite{HanngiFM} 
\begin{equation}
f(t) = e^{i\eta t}\phi(t), 
\end{equation}
and its complex conjugate, where $\phi(t)$ is a periodic function of
time with the same period as $\nu(t)$. $\eta$ is, in general, a complex
number \cite{WardFT}. The Floquet operators follow the usual
commutation relations for creation and annihilation operators
\begin{equation}\label{eq:floquet_com_rel}
[\Gamma^\dagger(t),\Gamma(t)]=1.
\end{equation}

Using these operators, $H_{\rm mec}(t)$ (see Eq.~\eqref{eq:Hmec}), can
be written in the same form as the non time-dependent harmonic
oscillator with the Floquet operators playing the role of the creation
and annihilation operators, with the exception of a global
time-dependent scalar coefficient \cite{BrownPT}

\begin{equation}\label{eq:floquet_only_h}
H_{\rm mec}(t) = \hbar\frac{W}{|f(t)|^2}\left[\Gamma^\dagger(t)\Gamma(t) + \frac{1}{2}\right],
\end{equation}
where $W$ is the Wronskian for the differential equation
\eqref{TimeDependentHO}. Using
Eqs.~(\ref{eq:floquet_com_rel}), (\ref{eq:floquet_only_h}) we can define
the number of excitations of a parametrically driven oscillator in a
manner analogous to the quantum harmonic oscillator: as the
expectation value of the number operator
$\expval{m}=\expval{\Gamma^\dagger\Gamma}$.

The explicit time dependence of the Floquet
operators will not be noted from now on for the sake of brevity.
Equation \eqref{FloquetOperators} can be inverted and solved for the
harmonic oscillator's position operator 

\begin{equation}
\hat{x} = \gamma'_+(t)\Gamma+\gamma'_-(t)\Gamma^\dagger,
\end{equation}
this expression can be substituted into the interaction Hamiltonian, getting
\begin{equation}
  H_{\rm int}(t) = g_c\sqrt{\frac{\hbar}{2M}}a^\dagger a[\gamma'_+(t)\Gamma +\gamma'_-(t)\Gamma^\dagger]\, .
\end{equation}

The explicit expressions for $\gamma_\pm(t)$ are obtained when
explicit expressions for the solutions $f(t)$ are available. The
Hamiltonian~\eqref{eq:Hfull} contains two separate harmonic
oscillator-like terms, $H_{\rm cav}$ and $H_{\rm mec}$, that commute.
This allow us to derive a master equation following the same procedure
depicted in \cite{HanngiFM} for the mechanical oscillator, and the
standard procedure for the cavity. This derivation involves the Markov
approximation; in previous attempts to study a parametrically driven
oscillator, the time dependence of the frequency was included after
the Markov approximation had been performed, via time-dependent ad-hoc
coefficients for the damping \cite{BarberisLC}. Under the formalism
developed in \cite{HanngiFM} the frequency's time dependence is
accounted for when the Markov approximation is performed, via the
Floquet operators. As demonstrated in \cite{HanngiFM}, the method
employed here is a more complete, and thus accurate, treatment.

The optomechanical master equation with improved dissipation model is
\begin{equation} \label{LCMasterEquation}
\dot{\rho} = \frac{1}{i\hbar}[H,\rho] +L_a\rho + L_\Gamma \rho,
\end{equation}
where
\begin{align}
L_a \rho =& - \frac{\kappa}{2}(n_p + 1)[a^\dagger a\rho + \rho a^\dagger a -2a\rho a^\dagger]  \\
 &- \frac{\kappa}{2}(n_p)[ aa^\dagger\rho + \rho  aa^\dagger -2a^\dagger\rho a]\, ,\nonumber
\end{align}

\begin{align}\label{eq:mechanical_dissipation}
  L_\Gamma \rho =& - \frac{\gamma}{2}(n_m + 1)[\Gamma^\dagger \Gamma\rho + \rho \Gamma^\dagger \Gamma -2\Gamma\rho \Gamma^\dagger]  \\
                 &- \frac{\gamma}{2}(n_m)[ \Gamma\Gamma^\dagger\rho + \rho  \Gamma\Gamma^\dagger -2\Gamma^\dagger\rho \Gamma]\, ,\nonumber
\end{align} 
$\kappa$ is the energy decay rate for the cavity, $\gamma$ is the
decay rate for the mechanical oscillator, $n_p$ is the number
  of thermal excitations of the bath at the frequency resonant with
  the cavity $\omega_{cav}$, and $n_m$ is known as the effective
  thermal-bath occupation number\cite{HanngiFM}. In the undriven case,
  $n_m$ reduces to the number of thermal excitation of the bath at the
  natural frequency of the mechanical oscillator $\nu$. In the absence
  of interaction between the cavity and the mechanical oscillator, the
  stationary state for the cavity is a thermal state with mean photon
  number $n_p$, and for the oscillator it is a thermal state with mean
  number $n_m$ of mechanical excitations.

The superoperators $L_\Gamma$ and $L_a$ model the
energy exchanges between the environment, and the cavity and the
mechanical resonator respectively.  Note that the time dependence
  of the Floquet operators implies that the dissipation of the
  mechanical oscillator, given by (\ref{eq:mechanical_dissipation}),
  is time-modulated. Equation \eqref{LCMasterEquation} is the
master equation for a parametrically driven optomechanical system with
an improved dissipation model which accounts for the mechanical
oscillator's time dependent frequency.

\subsection{Displaced Frame}

In order to eliminate the pump term and find useful approximations,
we employ a unitary transformation to shift equation \eqref{LCMasterEquation}
into a displaced reference frame. This transformation depends on two
time-dependent coefficients, $\alpha(t)$ and $\beta(t)$, which are
chosen in a convenient manner to simplify the Hamiltonian. The
transformation is given by the operator
\begin{equation}\label{ShiftTransform}
U_{a,\Gamma} = e^{(\alpha(t) a^\dagger - \alpha^*(t)a)}e^{(\beta(t) \Gamma^\dagger - \beta^*(t)\Gamma)},
\end{equation}
and results in a displaced Hamiltonian and in turn a displaced master
equation for the time evolution of the density operator
$\rho'(t)=U\rho U^\dagger$
\begin{equation}\label{eq:master_no_small}
\dot{\rho}' = \frac{1}{i\hbar}[H',\rho'] +L_a\rho' + L_\Gamma \rho' + C(t)\rho' ,
\end{equation}
where 
\begin{equation}
C(t)=-(\beta^2-|\beta|^2)[\dot{\Gamma}^\dagger,\Gamma^\dagger]-((\beta^*)^2-|\beta|^2)[\dot{\Gamma},\Gamma]\, .
\end{equation}
This term arises due to the explicit time dependence in the Floquet
operators as they do not, in general, commute with their own time
derivatives. The primes indicate that the transformation has been
applied. The displaced Hamiltonian, which includes a pump-like term
that appears due to the transformation being applied to the time
derivative term, is
\begin{align}\label{eq:hamiltonian_no_small}
  H'=&U H U^\dagger\nonumber\\=& -\hbar \delta' a^\dagger a + \hbar\frac{W}{|f(t)|^2}\Gamma^\dagger \Gamma\nonumber\\
     &-\hbar g_c\sqrt{\frac{\hbar}{2M}}[(a^{\dagger}a +\alpha a^{\dagger}+\alpha^* a)(\gamma'_-(t)\Gamma^{\dagger}+\gamma'_+(t)\Gamma)]\nonumber\\
     &+ i\hbar(\beta^*\dot{\Gamma} - \beta \dot{\Gamma}^\dagger),
\end{align}
with $\delta' = \delta + g_c\sqrt{\frac{\hbar}{2M}}(\gamma'_+(t)\beta + \gamma'_-(t)\beta^*)$.
This Hamiltonian is valid as long as the coefficients $\alpha(t)$ and
$\beta(t)$ fulfill the differential equations

\begin{align}\label{eq:displaced_frame}
\dot{\alpha} =& \alpha \Big(-\frac{\kappa}{2}+i(\delta+g_c\sqrt{\frac{\hbar}{2M}}(\gamma'_-(t) \beta^* + \gamma_+(t) \beta)\Big)-i\frac{\Omega}{2},\\
\dot{\beta} =& \beta \Big(-\frac{\gamma}{2}-i\frac{W}{|f(t)|^2}\Big)+ig_c\sqrt{\frac{\hbar}{2M}}|\alpha|^2\gamma'_-(t).
\end{align}

Proceeding further requires an explicit solution for equation
\eqref{TimeDependentHO} in order to calculate explicit expressions for several of the coefficients and to deal with the $\dot{\Gamma}$ and $\dot{\Gamma}^\dagger$ operators. The primes in the operators will be omitted
from now on as all calculations will be carried out in the displaced frame.

\section{Solution for Small Oscillations}\label{SolSmallOsc}
 
In order to obtain an explicit form of the Floquet operators we focus
on the case of small oscillations around a central frequency,
specifically

\begin{equation}
\nu(t) = \nu_0 + \epsilon' cos(2\omega t)\, ,
\end{equation}
with $\epsilon' \ll \nu_0$, where $\nu_0$ is the mean frequency. This
leads to the time-dependent harmonic oscillator equation
\begin{equation}\label{SmallOscillationsTDHO}
\ddot{f} + (\nu_0^2 + 2\epsilon' \nu_0 cos(2\omega t))f = 0,
\end{equation}
when neglecting terms of order 2 or higher in $\epsilon$ and
represents a particular case of the Mathieu equation~\cite{PiatekME}.
In order to guarantee stable solutions with the required
periodicity~\cite{WardFT} we require the scattering relation
\begin{equation}
\frac{\nu_0^2}{\omega^2} = n^2,\label{scattering}
\end{equation}
with $n \in \mathbb{Z}^+$. The solutions for equation
\eqref{SmallOscillationsTDHO} are, to first order in
$\epsilon= \frac{2\epsilon' \nu_0}{\omega^2}$,
\begin{align}\label{SmallOscillationsSolution}
f(t)=  \frac{1}{\sqrt{n\omega}}\big(e^{in\omega t}  &+ \epsilon \frac{1}{8(n+1)} e^{i(n+2) \omega t}\nonumber\\
&- \epsilon \frac{1}{8(n-1)} e^{i(n-2) \omega t}\big),
\end{align}
and its complex conjugate. To simplify the comparison with the
non-parametrically driven case we define
\begin{align}
    \Tilde{\Gamma}(t)=&e^{-in\omega t}\Gamma(t),\\
    \Tilde{\Gamma}^\dagger(t)=&e^{in\omega t}\Gamma^\dagger(t).
\end{align}
These operators retain the same commutation relations as the original
$\Gamma$ operators. In general, any term involving the same number of
$\Gamma$ and $\Gamma^\dagger$ operators is unchanged. All calculations
will use the $\Tilde{\Gamma}$ operators, so the tilde will be omitted
from this point. The operators can be written as

\begin{align}\label{eq:GammaOperatorTimeAsCorrection}
\Gamma(t)=&\frac{1}{2i}\left[\hat{x}\sqrt{\frac{2M}{\hbar}}h(t)-\hat{p}\sqrt{\frac{2}{M \hbar }}g(t)\right],\\
\Gamma^\dagger(t)=&\frac{-1}{2i}\left[\hat{x}\sqrt{\frac{2M}{\hbar}}h^*(t)-\hat{p}\sqrt{\frac{2}{M \hbar }}g^*(t)\right]
\end{align} with

\begin{align*}
g(t)=&\frac{1}{\sqrt{n\omega}}(1+\frac{\epsilon}{8(n+1)}e^{2i\omega t}-\frac{\epsilon}{8(n-1)}e^{-2i\omega t}),\\
h(t)=&\frac{1}{\sqrt{n\omega}}(in\omega+\frac{\epsilon i\omega (n+2)}{8(n+1)}e^{2i\omega t}-\frac{\epsilon i\omega (n-2)}{8(n-1)}e^{-2i\omega t}).
\end{align*}
We can then calculate all of the time dependent terms that require
specific solutions for $f(t)$, which can be just as easily obtained in
terms of $g(t)$ and $h(t)$. We have, neglecting terms of order
$\frac{\epsilon}{n^2}$

\begin{equation}
    \frac{W}{\abs{f(t)}} \approx \nu_0,
\end{equation}

and

\begin{align}
    \gamma_+(t) =&g^*(t)\, ,\\
    \gamma_-(t) =&g(t)\, .
\end{align}
With these coefficients we can solve equations
\eqref{eq:displaced_frame}. We are interested in case where the
stationary case is reached on a small time scale. In that case we can
assume that $\dot{\alpha}=\dot{\beta}=0$. We also assume that the
coupling is weak enough to be neglected at first order. The solutions
are then

\begin{align}\label{eq:AlfaBetaSolutions}
    \alpha_0 =& \frac{\Omega}{2\delta + i\kappa},\\
    \beta_0 =& 0.
\end{align}
The subscript indicates that these solutions are valid up to order
$0$ in the coupling parameter.
The Hamiltonian is now

\begin{equation}\label{eq:hamiltonian_final}
  H= -\hbar \delta a^\dagger a + \hbar\nu_0\Gamma^\dagger \Gamma-H_{int},
\end{equation} with

\begin{equation*}
H_{int}(t) = g_c\sqrt{\frac{\hbar}{2M}}(\alpha_0^* a + \alpha_0 a^\dagger)(g^*(t)\Gamma(t) + g(t)\Gamma^\dagger (t)).
\end{equation*}
Setting $\chi_0 = g_c\sqrt{\frac{\hbar}{\nu_0 2 M}}$ we write
\begin{displaymath}
  H_{int}(t) = H_{int}^0 + H_{int}^\epsilon(t)\, ,
\end{displaymath}  
with
\begin{align*}
H_{int}^0(t) = &\chi_0(\alpha_0^* a + \alpha_0 a^\dagger)(\Gamma(t) + \Gamma^\dagger (t)) \, ,\\
H_{int}^\epsilon(t)=&\chi_0\epsilon(\alpha_0^* a + \alpha_0 a^\dagger)\times\\&\Big(\frac{1}{8(n+1)}(e^{-2i\omega t}\Gamma(t) + e^{2i\omega t}\Gamma^\dagger (t))\\
& -\frac{1}{8(n-1)}(e^{2i\omega t}\Gamma(t) + e^{-2i\omega t}\Gamma^\dagger (t))\Big ).
\end{align*} Due to \eqref{eq:AlfaBetaSolutions}, the terms in the master
equation involving the time derivatives of the Floquet operators, both
the commutator terms and the pump-like term, vanish. Then, the master
equation (\ref{eq:master_no_small}) can be written as
\begin{equation}\label{eq:LCMasterEq}
\dot{\rho} = \frac{1}{i\hbar}[H,\rho] +L_a\rho + L_\Gamma \rho = \mathcal{L}\rho.
\end{equation}
This last equation is a model for a mechanical oscillator with time
dependent frequency, interacting with an electromagnetic field, and
with a dissipation model that takes into account that the mechanical
object's frequency depends on time. It looks similar to the standard
optomechanical master equation but it has Floquet operators instead of
creation and annihilation operators for the mechanical oscillator and
an explicit time dependence in the interaction Hamiltonian. It is one
of the main results of this paper, it gives the evolution of the
parametrically driven optomechanical system with an improved
dissipative model. In the next sections we will focus on calculating
the number of excitations of the mechanical object
$\expval{m}=\expval{\Gamma^\dagger\Gamma}$.

\section{Laser Cooling}\label{LasCool}

We use the master equation \eqref{eq:LCMasterEq} to study laser
cooling of the parametrically driven mechanical object. Our goal is to
minimize the temperature of the mechanical object. In the displaced
frame, where the master equation is written,
this is equivalent to minimizing the number of mechanical excitations.
Our focus is on the parameter regime where the coupling is weak enough
that we may take $\alpha_0$ and $\beta_0$ to be the solutions to
\eqref{eq:displaced_frame}. After projecting into the subspace
corresponding to the slowly evolving time scale and tracing over the
cavity degrees of freedom, we arrive at the following master equation
for the density operator $\mu(t) = Tr_c[P\rho(t)]$, 
\begin{equation}\label{eq:ProyectedMasterEqCoolingGamma}
\dot{\mu} = (A_-(t)+\frac{\gamma}{2}(n_m+1))D[\Gamma]\mu + (A_+(t)+\frac{\gamma}{2}n_m)D[\Gamma^\dagger]\mu, 
\end{equation}
with
  $D[\Gamma] = 2\Gamma \mu \Gamma^\dagger -\{\Gamma^\dagger \Gamma,
  \mu\}$. $A_-(t)$ and $A_+(t)$ are known as the cooling and heating rates,
  respectively. 
  This equation is obtained in Appendix
  \ref{CoolingAppendix}.
The coefficients
\begin{equation}\label{eq:ACoefficients}
    A_{\pm}(t) = A^0_{\pm} + \epsilon \sin(2\omega t) A^\epsilon_{\pm}\, ,
\end{equation}
can be written as the usual rates for the non driven case
\begin{equation*}
A^0_\pm=\frac{\chi_0^2 \abs{\alpha_0}^2}{2}\frac{\kappa}{(\delta\mp
  \nu_0)^2+\frac{\kappa^2}{4}}\, ,
\end{equation*}
plus a correction proportional to $\epsilon$
\begin{equation*}
A_{\pm}^\epsilon= \frac{\chi_0^2
  \abs{\alpha_0}^2}{2}\frac{(\delta\mp\nu_0) }{n\left(
    \frac{{\kappa}^{2}}{4}+{{(\delta\mp\nu_0)}^{2}}\right)}\, .
\end{equation*}
We wish to obtain an expression for the mean number of mechanical
excitations $\expval{m}$,  which is a measure for the system's temperature. We use
the system's covariance matrix~\cite{ZollerCovMat} to do that. Defining
\begin{align}
    X =& \sqrt{\frac{n\omega}{2}}(g(t)^* \Gamma + g(t) \Gamma^\dagger),\\
    P =& \frac{1}{\sqrt{2n\omega}}(h(t)^* \Gamma + h(t) \Gamma^\dagger),
\end{align}
and 
\begin{equation}
    \overline{R} = [X,P]^T,
  \end{equation}
the expectation value of the covariance matrix is then expressed as
\begin{equation}
    \overline{\gamma}_{i,j} =
   \frac{1}{2}\expval{\overline{R}_i\overline{R}_j+\overline{R}_j\overline{R}_i}-\expval{\overline{R}_i}\expval{\overline{R}_j}\,.
\end{equation}
The calculations to obtain an expression for the covariance matrix
$\overline{\gamma}(t)$ are performed in Appendix \ref{app:Covariance}.

The mean number
\begin{equation}\label{eq:number_mechanical}
    \expval{m} = \frac{1}{2}(Tr[\overline{\gamma}]-1)\, ,
\end{equation} 
of mechanical excitations, can be calculated as a
function of the trace of $\overline{\gamma}$ \cite{BarberisLC}.
Defining
\begin{eqnarray*}
  \tilde{A}_-^0 &=& A_-^0+\frac{\gamma}{2}(n_m+1)\, ,\\
  \tilde{A}_+^0 &=& A_+^0+\frac{\gamma}{2}n_m\, ,
\end{eqnarray*}
we obtain that
\begin{align}\label{eq:CovarianceMatrixTrace}
    Tr[\overline{\gamma(t)}] =& \frac{\tilde{A}_+^0+\tilde{A}_-^0}{\tilde{A}_-^0-\tilde{A}_+^0}\\
    &-\epsilon(A_+^\epsilon+A_-^\epsilon)\frac{(\tilde{A}_+^0-\tilde{A}_-^0)\sin(2\omega t)}{(\tilde{A}_+^0-\tilde{A}_-^0)^2+\omega^2}\nonumber\\
    &-\epsilon(A_+^\epsilon+A_-^\epsilon)\frac{\omega\cos(2\omega t)}{(\tilde{A}_+^0-\tilde{A}_-^0)^2+\omega^2}\nonumber\\
    &+\frac{\epsilon}{\omega}\frac{(A_+^\epsilon-A_-^\epsilon)(\tilde{A}_+^0-\tilde{A}_-^0)(\tilde{A}_+^0+\tilde{A}_-^0)}{(\tilde{A}_+^0-\tilde{A}_-^0)^2+\omega^2}\nonumber\\
    &-\frac{\epsilon}{\omega}\frac{(A_+^\epsilon - A_-^\epsilon)(\tilde{A}_-^0+\tilde{A}_+^0)}{(\tilde{A}_+^0-\tilde{A}_-^0)}\nonumber.
\end{align}

These results are valid when $\delta<0$, if $\delta>0$ we have heating
and the number of mechanical excitations diverges in the framework of
our theory. The number of mechanical excitations,
Eq.~(\ref{eq:number_mechanical}), has four correction terms
proportional to $\epsilon$, two are time independent and the other two
oscillate with the frequency of the drive. As expected, when
$\omega\rightarrow 0$, $\expval{m}$ becomes the number of mechanical
excitations for the non parametrically driven case. To first order in
$\omega$, the time independent corrections vanish, and the number of
mechanical excitations oscillates around the non parametrically driven
case with a frequency given by $2\omega$. If we take the time average
of $\expval{m}$ over one period, there will be a time independent
correction of order $\omega^2$. We now analyze the effects of the
correction terms on the number of mechanical excitations. In the
unresolved sideband regime ($\kappa>>\nu_0$), the constant correction
terms tend towards zero (as in this regime
$A^\epsilon_+ \approx A^\epsilon_-$). The time dependent corrections
remain, but these are zero when time averaged. Note that
$A^0_\pm\sim\frac{1}{\kappa}$ and
$A^\epsilon_\pm\sim \frac{1}{\kappa^2}$ so as $\kappa$ increases, the
correction terms become irrelevant.

We will focus in the resolved sideband regime $\kappa<\nu_0$. 

 When
$\omega^2\ll (A_+^0-A_-^0)^2$, we can approximate the average, over
one period of time, of the number of mechanical excitations
\begin{equation*}
\expval{\overline{m}}=\frac{\pi}{\omega}\int_0^{\pi/\omega}\expval{m}dt\, ,
\end{equation*}
as
\begin{equation}
  \label{eq:approx_nm}
  \expval{\overline{m}}\approx
  \expval{\overline{m}}_{n_m=0}+\frac{\gamma
    n_m}{\Gamma_{\rm cool}+\gamma/2}+\epsilon\omega\frac{\gamma n_m}{2(\Gamma_{\rm cool}+\gamma/2)^3}(A_+^\epsilon
  - A_-^\epsilon)\, ,
\end{equation}
where $\Gamma_{\rm cool}=A_-^0-A_+^0$ and
\begin{equation}
  \label{eq:mnm0}
  \expval{\overline{m}}_{n_m=0}\approx \frac{A_+^0}{\Gamma_{\rm cool}+\gamma/2}+\epsilon\omega\frac{A_-^0+A_+^0+\gamma/2}{2(\Gamma_{\rm cool}+\gamma/2)^3}(A_+^\epsilon
  - A_-^\epsilon)\, .
\end{equation}
The second and third term in Eq.~(\ref{eq:approx_nm}) are the
contribution, to the mean number of mechanical excitations, when the
temperature of the mechanical bath is not zero. When
$\Gamma_{\rm cool}\gg\gamma n_m$ this contribution is negligible and we obtain
$\expval{\overline{m}}\approx \expval{\overline{m}}_{n_m=0}$.

\begin{figure}
    \centering
    \includegraphics[scale=.45]{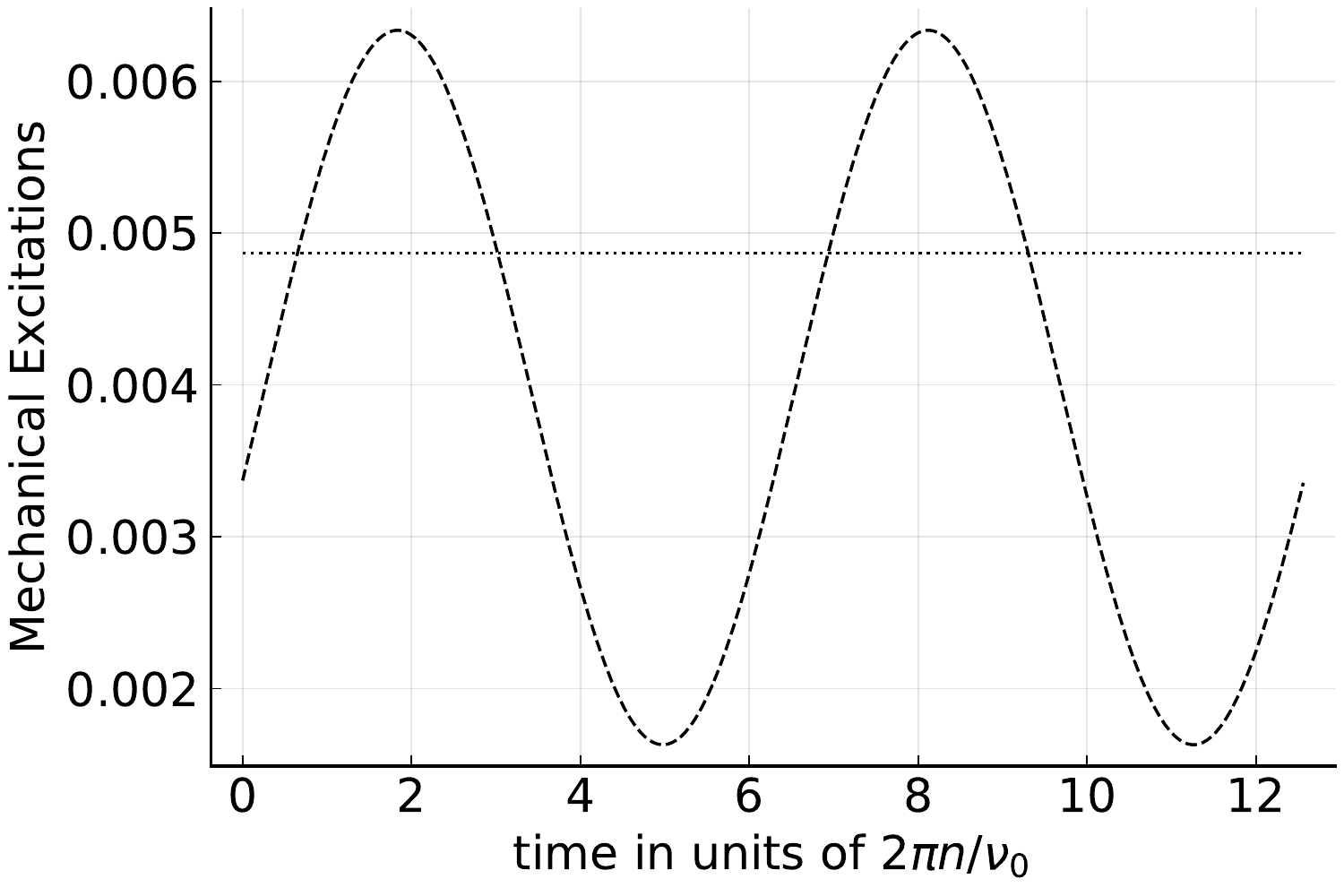}
    \caption{Number of mechanical excitations for the parametrically (dashed line)
      and non parametrically (dotted line) driven case. The number of mechanical
      excitations oscillates with the frequency given by the drive
      and is smaller than in the non parametrically driven case for
      a majority of the time period.
      Parameters: $\delta/\nu_0=-0.9469$, $n=2$,
      $\epsilon=1/18$, $\kappa=0.25 \nu_0$,
      $\chi_0^2\abs{\alpha_0}^2/\nu_0^2=0.25$.}
    \label{fig:TDetuningFixed}
\end{figure}

Sideband cooling is used as a final cooling stage \cite{ParkSidebandCryogenic2009}.  When sideband cooling is begun, $n_m$ can be in the range of 1000
  excitations and $\Gamma_{\rm cool}\gg\gamma n_m$
  \cite{PetersonMicromechanicalMembraneBackactionLimit2016}.
  Under
  these conditions we have that $\tilde{A}_\pm^0\approx A_\pm^0$ and
  we get that 
\begin{widetext}
\begin{eqnarray}
  \label{eq:aprox_to_aprox}
  \expval{\overline{m}}&\approx&-\frac{(\nu_0+\delta)^2+\kappa^2/4}{4\delta\nu_0}+\frac{\epsilon\omega}{32\delta^3\kappa^2\nu_0^3\chi_0^2\abs{\alpha_0}^2}\left[(\nu_0^2-\delta^2+\kappa^2/4)(\nu_0^2+\delta^2+\kappa^2/4)((\nu_0+\delta)^2+\kappa^2/4)((\nu_0-\delta)^2+\kappa^2/4)\right]\,
                                 .\nonumber\\
\end{eqnarray}
\end{widetext}
When $\nu_0^2-\delta^2+\kappa^2/4=0$ there is no difference in the
number of mechanical excitations between the parametrically driven
and the non
parametrically driven cases. Note that when $\epsilon>0$ and
$\delta^2<\nu_0^2+\kappa^2/4$, or $\epsilon<0$ and
$\delta^2>\nu_0^2+\kappa^2/4$, the mean number of mechanical
excitations in the parametrically driven case is smaller than in the
non parametrically driven case. An example of this is shown in figure
\ref{fig:TDetuningFixed}. The detuning is chosen for the case where
$\expval{\overline{m}}$ is minimal at $t=0$ and the number of mechanical
excitations $\expval{m}$ is plotted, as a function of time, for the
non parametrically driven and the parametrically driven cases. The
number of mechanical excitations, for the parametrically driven case,
is lower than the non parametrically driven case for most of the time period.

\begin{figure}
    \centering
    \includegraphics[scale=0.5]{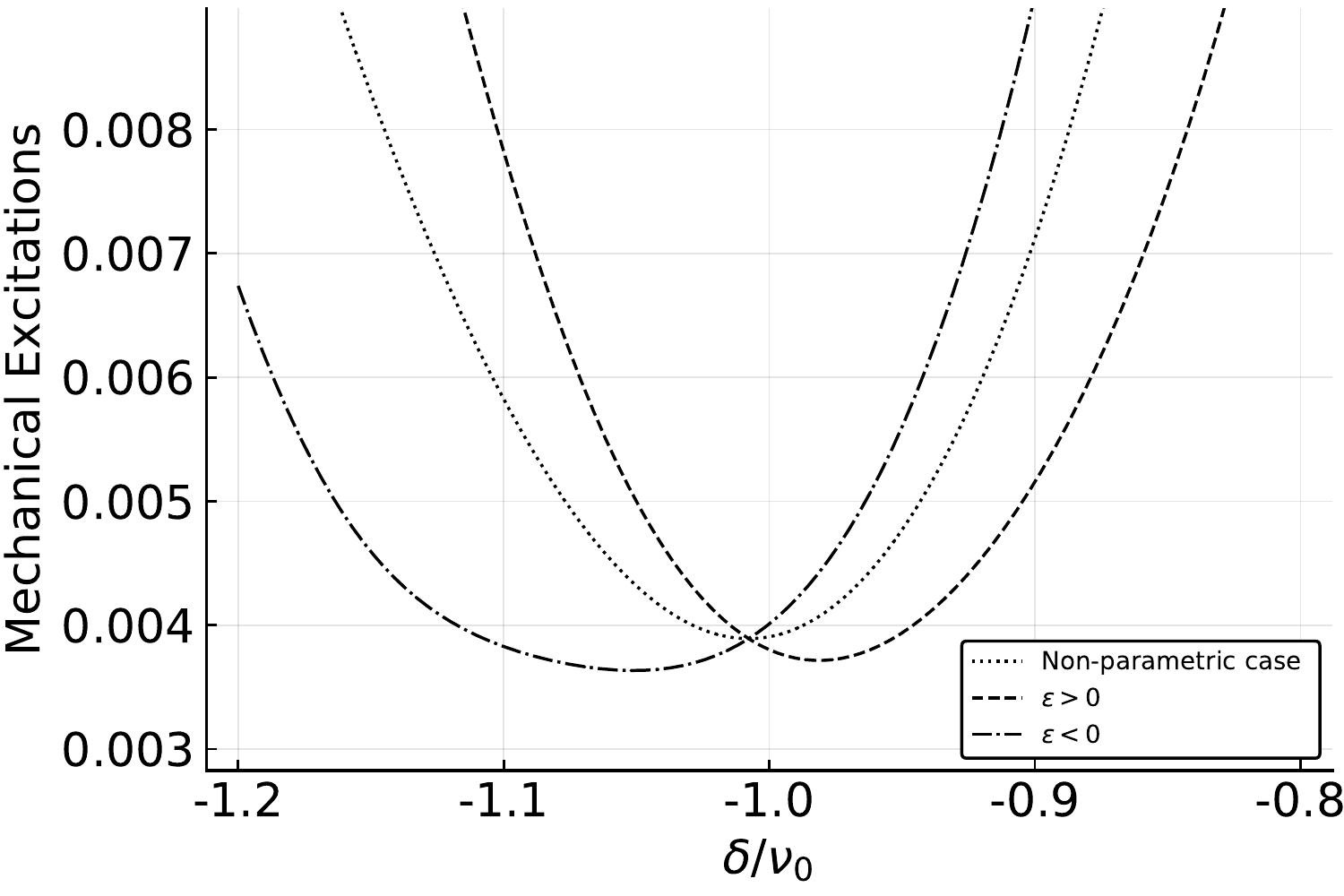}
    \caption{Time averaged number of mechanical excitations for the
      parametrically (dashed line for $\epsilon > 0$, dot-dash line for $\epsilon <0$) and non parametrically (dotted line) driven case as a function
      of detuning. Parameters: $n=2$,
      $\epsilon=\pm 1/18$, $\kappa=0.25 \nu_0$,
      $\chi_0^2\abs{\alpha_0}^2/\nu_0^2=0.25$.}
    \label{fig:TempVsDetuningInsetRatio}
\end{figure}

\begin{figure}
    \centering
    \includegraphics[scale=0.5]{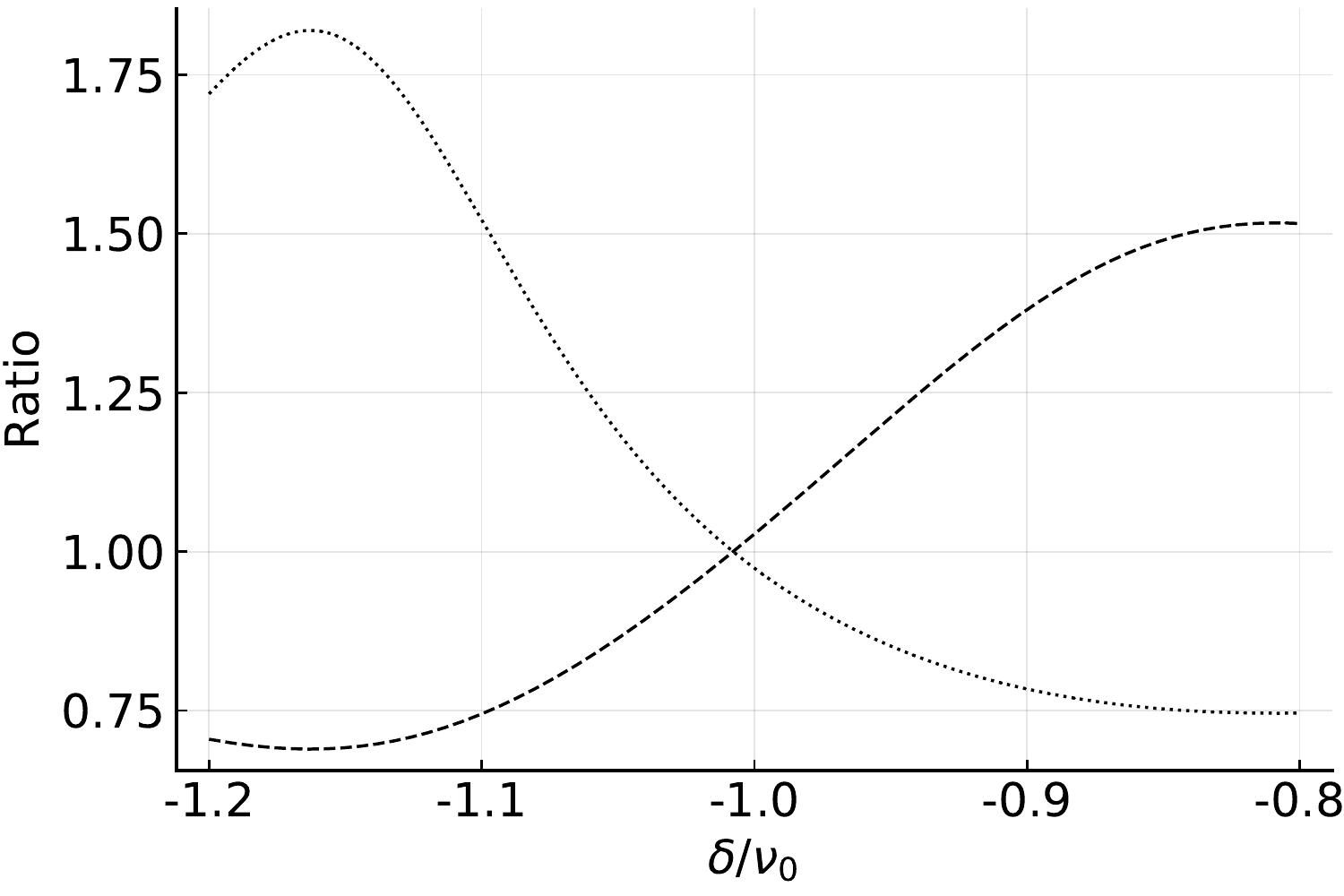}
    \caption{Ratio, as a function of detuning, between the time
      averaged number of mechanical excitations for the parametrically (dashed line for $\epsilon > 0$, dotted line for $\epsilon <0$)
      and the non parametrically driven case . Parameters: $n=2$,
      $\epsilon=\pm 1/18$, $\kappa=0.25 \nu_0$,
      $\chi_0^2\abs{\alpha_0}^2/\nu_0^2=0.25$}
    \label{fig:TempRatioVsDetuning}
\end{figure}

In some cases $\expval{\overline{m}}$ can be smaller than the smallest achievable
temperature in the non parametrically driven case. To show this we
calculate $\expval{\overline{m}}$, over the range
$\delta/\nu_0=[-1.2,-0.8]$, for the parametrically and non
parametrically driven case. The result can be seen in figure
\ref{fig:TempVsDetuningInsetRatio}. We can see that the minimum number
of mechanical excitations can be lower than in the non parametrically
driven case. The value of $\delta$, where the minimum is achieved,
depends on the sign of $\epsilon$, as predicted by
Eq.~(\ref{eq:aprox_to_aprox}). In Fig.~\ref{fig:TempRatioVsDetuning}
we compare the ratio of $\expval{\overline{m}}$ between the
parametrically and non parametrically driven cases. For the parameters
in the figure, which are consistent with the approximations used in
the calculations, the difference can be up to 10\%.

\section{Conclusions}\label{ConCl}

Using an improved theoretical description for the dissipation of a
parametrically driven mechanical object in an optomechanical setup, we
found that the temperature can be lower than in the non-driven case.
Moreover, the usage of a consistent dissipation model affects the
predictions for the cooling dynamics. The results of this paper allow
for the analysis of the discrepancy when compared to conventional
approaches. 
\begin{acknowledgments}
Support by project UNAM-PAPIIT IG100518.
\end{acknowledgments}
\appendix
\section{The Damping Basis}\label{App1}

Master equations of the type 

\begin{equation}
\dot{\rho} = \mathcal{L}_{cav} \rho = \frac{1}{i\hbar}[H,\rho]+L_a\rho, 
\end{equation} with

\begin{align}\label{EMField}
L_a \rho =& - \frac{\kappa}{2}(n_p+1)[a^\dagger a\rho + \rho a^\dagger a -2a\rho a^\dagger] \nonumber \\
 &- \frac{\kappa}{2}(n_p)[ aa^\dagger\rho + \rho  aa^\dagger -2a^\dagger\rho a]\, ,
\end{align}
and
\begin{equation}
  H=\hbar \omega_c \, a^\dagger a\, ,
\end{equation}
model the behavior of a bosonic field inside a one mode leaky cavity
with frequency $\omega_c$; the cavity is in contact with a thermal bath
characterized by $n_p$ thermal photons, the cavity damping is given by
$\kappa$ \cite{EnglertDB} and $a^\dagger$, $a$ are cavity photons
creation and anhilation operators. The density operator can be
expressed in a basis given by the right Lindblad superoperator's
eigenstates, ${\hat{\rho}_n^j}$,
$n=0,1,2,...\qquad j = 0,\pm 1, \pm 2,... $, where
\begin{equation}
L\hat{\rho}_n^j = \lambda_n^j\hat{\rho}_n^j\label{eq:eigen_damping}\, ,
\end{equation}
with
\begin{equation}
\lambda_n^j = ij\omega_c -\kappa[n + \frac{|j|}{2}]\, .
\end{equation}
The real part of these eigenvalues corresponds to the 
eigenvalues of the operator $L_a$. This basis is known as the damping
basis \cite{EnglertDB} and is given by
\begin{align}\label{DefDB}
\hat{\rho}_n^l=&a^{\dagger j}\frac{(-1)^n}{(n_p+1)^{j+1}}:L_n^l[\frac{a^\dagger a}{n_p+1}]e^{-[\frac{a^\dagger a}{n_p+1}]}:\quad j \geq 0, \\
\hat{\rho}_n^j=&\frac{(-1)^n}{(n_p+1)^{|j|+1}}:L_n^{|j|}[\frac{a^\dagger a}{n_p+1}]e^{-[\frac{a^\dagger a}{n_p+1}]}:a^{|j|}\quad j \leq 0.
\end{align}
The Lindblad operator is not hermitian and the left eigenstates must
be considered to find the coefficients of the density operator
expansion in the damping basis. These are the eigenstates of the
equation $\check{\rho}_n^jL = \lambda_n^j\check{\rho}_n^j$ they have
the same eigenvalues and are given by
\begin{align}\label{DefDBDual}
\check{\rho}_n^j=&(\frac{-n_p}{n_p+1})^n\frac{n!}{(n+j)!}:L_n^j[\frac{a^\dagger a}{n_p}]:a^{j}\quad j \geq 0, \\
\check{\rho}_n^j=&(\frac{-n_p}{n_p+1})^n\frac{n!}{(n+|j|)!}a^{\dagger|j|}:L_n^{|j|}[\frac{a^\dagger a}{n_p}]:\quad j \leq 0.
\end{align}

The left and right eigenstates are orthogonal under the product
\begin{equation}
(\hat{\rho}_n^j,\check{\rho}_{n'}^{j'})=Tr[\hat{\rho}_n^j\check{\rho}_{n'}^{j'}] = \delta_{n,n'}\delta_{j,j'},
\end{equation} and fulfill

\begin{equation}\label{DampingBasisCompleteness}
\sum_{\lambda} \hat{\rho}_\lambda \otimes \check{\rho}_\lambda = \mathbb{I},
\end{equation} where the sum is over all possible eigenvalues. An important case is a cavity at zero temperature, in this case the right states are \cite{EnglertDB}

\begin{align}\label{DefDBZero}
\hat{\rho}_n^j=&a^{\dagger j}(-1)^{a^\dagger a + n}\binom{n+j}{a^\dagger a+j} \quad j \geq 0, \\
\hat{\rho}_n^j=&(-1)^{a^\dagger a + n}\binom{n+|j|}{a^\dagger a+|j|}a^{|j|} \quad j < 0,
\end{align} and the left states are

\begin{align}\label{DefDBDualZero}
\check{\rho}_n^j=&\frac{n!}{(n+j)!}\binom{a^\dagger a}{n}a^j \quad j \geq 0, \\
\check{\rho}_n^j=&a^{\dagger|j|}\frac{n!}{(n+|j|)!}\binom{a^\dagger a}{n} \quad j < 0.
\end{align}

These states play an important part in the derivation of the master
equation of the cavity state in the adiabatic approximation.

In Appendix \ref{CoolingAppendix} we consider a harmonic oscillator
with no damping, so the left and right eigenstates of the damping
basis reduce to

\begin{equation}
\hat{\rho}_{n}^{l}=\Ket{n+l}\Bra{n}=\check{\rho}_{n}^{\dagger l},
\end{equation}
with eigenvalues
\begin{equation}
\lambda_l = i l \nu_0\, ,
\end{equation}
$\Ket{n}$ is the number state of the harmonic oscillator, $l$ is an
integer satisfying $(n+l)>0$.

\section{Laser Cooling and Projection Operators}\label{CoolingAppendix}

In order to find the master equation
\eqref{eq:ProyectedMasterEqCoolingGamma} we begin with the equation based on the Hamiltonian \eqref{eq:hamiltonian_final}
\begin{equation}\label{eq:master_no_mechanical_damping}
\dot{\rho}=(\mathcal{L}_0+\mathcal{L}_1)\rho\, ,
\end{equation}
where
\begin{eqnarray}\label{eq:freelindblat}
\mathcal{L}_0=&\mathcal{L}_{cav} + \mathcal{L}_{mec},\\
 =&(\frac{1}{i\hbar}[H_{\rm cav},\bullet]+L_a) +(\frac{1}{i\hbar}[H_{\rm mec},\bullet]), \nonumber
\end{eqnarray}
gives the free dynamics 
and
\begin{equation}
\mathcal{L}_1 =\mathcal{L}_1^0+\mathcal{L}_1^\epsilon= \frac{1}{i\hbar}[H^0_{int}+H^\epsilon_{int},\bullet],
\end{equation}
gives the field-mechanic oscillator interaction.

Equation \eqref{eq:master_no_mechanical_damping} is the same as
equation \eqref{eq:master_no_small} without the mechanical damping, which
occurs on a slower time scale than the other processes and can be
incorporated after the adiabatic approximation. We employ projection
operators, like those in \cite{CarmichaelQO}, to separate the
evolution into different time scales and perform an adiabatic
approximation. The projection operator $P$ projects the state into a
slow-decaying evolution space whereas the projection operator $Q$
projects the system into a fast-decaying evolution space, the projection operators fulfill
the completeness relation
\begin{equation}
1 = P + Q,
\end{equation}
and have the properties
\begin{enumerate}

\item $ P\mathcal{L}_{0} = \mathcal{L}_{0}P = 0 $\qquad as P projects the state to the stationary subspace
\item $P\mathcal{L}_{1}P=0$ \qquad as the interaction does not couple states in P

\item $P^2 = P \quad Q^2 = Q$ \qquad as $P$ and $Q$ are projectors.
\end{enumerate}
In the decay picture the master equation is 
\begin{equation}
  \label{eq:master_decay_picture}
    \dot{\rho'} = \mathcal{L}'_1\rho'\, ,
\end{equation}
where
\begin{align*}
 \rho' =& e^{\int_0^t \mathcal{L}_0 dt'}\rho,\\
  \mathcal{L}_1' =& e^{-\int_0^t \mathcal{L}_0 dt'}\mathcal{L}_1e^{\int_0^t \mathcal{L}_0 dt'}.
\end{align*} or more explicitly

\begin{align}
\rho' =& e^{-\mathcal{L}_0t}\rho,\\
\mathcal{L}_1' =&e^{-\mathcal{L}_0t}\mathcal{L}_1' e^{\mathcal{L}_0t}.
\end{align} We project the master equation
\eqref{eq:master_decay_picture} into both $P$ and $Q$ to obtain the
equations
\begin{align*}
P\dot{\rho}' =& P\mathcal{L}_1'Q\rho', \\
Q\dot{\rho}' =& Q\mathcal{L}'_1Q\rho' + Q\mathcal{L}'_1P\rho'.
\end{align*} The equation for $Q$ can be formally integrated

\begin{align*}
Q\rho =& Q\rho'(t_0) + \int_{t_0}^{t}dt' Q\mathcal{L}'_1(t')P\rho'(t')\\
       &+\int_{t_0}^{t}dt'Q\mathcal{L}'_1(t')Q\rho'(t'),
\end{align*}
 and then the Markov approximation is performed, approximating $\rho(t')$ by $\rho(t_0)$
\begin{align*}
Q\rho\simeq & Q\rho'(t_0) + \int_{t_0}^{t}dt' Q\mathcal{L}'_1(t')P\rho'(t_0)\\
       &+\int_{t_0}^{t}dt'Q\mathcal{L}'_1(t')Q\rho'(t_0),
\end{align*}
and this is substituted into the $P$ equation
\begin{align}
P\dot{\rho'}(t) =& P\mathcal{L}_1Q\rho'(t_0)\\ 
 &+ P\mathcal{L}_1\int_{t_0}^{t}dt' Q\mathcal{L}_1(t')P\rho'(t_0)\nonumber \\
 &+ P\mathcal{L}_1\int_{t_0}^{t}dt'Q\mathcal{L}_1(t')Q\rho'(t_0)\nonumber,
\end{align}
where only the second term is non zero as we can choose the initial
condition to have no part in $Q$. We focus on this term and transform
back from the decay picture

\begin{align}\label{eq:temp_1}
P\dot{\rho}'(t)=&P e^{-\mathcal{L}_0 t}\mathcal{L}_1e^{\mathcal{L}_0 t}\\
&\int_{t_0}^{t}dt'Qe^{-\mathcal{L}_0 t'}\mathcal{L}_1e^{\mathcal{L}_0 t'}Pe^{-\mathcal{L}_0 t_0}\rho(t_0).\nonumber
\end{align} We write the projectors as
\begin{eqnarray}
  P &=& \sum_{\lambda} (\hat{\rho}_{\lambda}^{cav}\otimes\hat{\rho}_{\lambda}^{mec})\otimes(\check{\rho}_{\lambda}^{cav}\otimes\check{\rho}_{\lambda}^{mec}),\label{eq:projector_p}\\
  &=& \sum_\lambda \mathcal{P}_\lambda, \nonumber\\
  Q &=& \sum_{\lambda'} (\hat{\rho}_{\lambda'}^{cav}\otimes \hat{\rho}_{\lambda'}^{mec})\otimes(\check{\rho}_{\lambda'}^{cav}\otimes\check{\rho}_{\lambda'}^{mec})\label{eq:projector_q}\, ,\\
  &=& \sum_\lambda \mathcal{Q}_{\lambda'}, \nonumber
\end{eqnarray}
the projectors with the $\lambda$ label project the state into the
slow-decaying time-scale subspace, they are eigenstates of
$\mathcal{L}_0$ with only eigenvalues equal to zero. The projectors with the
$\lambda'$ label corresponds to the fast-decaying time-scale, they are
eigenstates of $\mathcal{L}_0$ with eigenvalues with a non-zero real
part, those states decay quickly. The projectors are applied via the product

\begin{equation}
PX =\sum_\lambda \hat{\rho_{\lambda}}Tr[\check{\rho}_{\lambda}X],
\end{equation} with

\begin{equation}
\hat{\rho}_\lambda = \hat{\rho}_\lambda^{mec} \otimes  \hat{\rho}_\lambda^{cav}.
\end{equation} We employ the states defined in Appendix
\ref{App1} for both the cavity and the mechanical resonator. 

Using equations \eqref{eq:projector_p} and \eqref{eq:projector_q} in
\eqref{eq:temp_1} and applying the operator $\mathcal{L}_0$ we obtain
\begin{align}\label{ProyectionEQ}
P\dot{\rho}'(t)=P e^{-\mathcal{L}_0 t}\mathcal{L}_1\Big(&\sum_{\lambda',\lambda}\int_{t_0}^{t}dt'e^{\lambda't} \hat{\rho}_{\lambda'} \otimes \check{\rho}_{\lambda'}e^{-\lambda't'}\mathcal{L}_1\\
&e^{\lambda t'}\hat{\rho}_{\lambda} \otimes \check{\rho}_{\lambda} e^{-\lambda t_0}\rho(t_0)\Big)\nonumber.
\end{align}
$\mathcal{L}_1$ is time-independent and the integration can be easily
performed. Returning $P$ and $Q$ to their original notation we may
write

\begin{align}
P\dot{\rho}'(t)=P e^{-\mathcal{L}_0 t}\mathcal{L}_1\Big(&\sum_{\lambda',\lambda}e^{\lambda't-\lambda t_0}\\
&\int_{t_0}^{t}dt' \mathcal{Q}_{\lambda'}e^{(\lambda-\lambda')t'}\mathcal{L}_1 \mathcal{P}_\lambda\rho(t_0)\Big)\nonumber.
\end{align} The integration is straightforward and we obtain

\begin{align}
P\dot{\rho}'(t)=P e^{-\mathcal{L}_0 t}\mathcal{L}_1\Big(&\sum_{\lambda',\lambda}\frac{1}{(\lambda-\lambda')}e^{\lambda't-\lambda t_0}\\
& \mathcal{Q}_{\lambda'}(e^{lt}-e^{lt_0})\mathcal{L}_1\mathcal{P}_\lambda\rho(t_0)\Big)\nonumber,
\end{align} which, after multiplying out exponentials within the sum results in

\begin{align}
P\dot{\rho}'(t)=P e^{-\mathcal{L}_0 t}\mathcal{L}_1\Big(&\sum_{\lambda',\lambda}\frac{1}{(\lambda-\lambda')}(e^{\lambda(t-t_0)}-e^{\lambda'(t-t_0)})\\
& \mathcal{Q}_{\lambda'}\mathcal{L}_1 \mathcal{P}_\lambda\rho(t_0)\Big)\nonumber.
\end{align}
We neglect terms proportional to $e^{\lambda' t}$ because for the slow
time scale these terms tend to zero. Using that $\lambda=0$, we can write

\begin{align}
    P\dot{\rho}(t)=\sum_{\lambda'}&\Big(\frac{-1}{\lambda'} P\mathcal{L}_1^0(t)\mathcal{Q}_{\lambda'}\mathcal{L}_1^0(t)P\rho(0)\nonumber\\ 
    &-\frac{1}{\lambda'} P\mathcal{L}_1^0(t)\mathcal{Q}_{\lambda'}\mathcal{L}_1^\epsilon(t)P\rho(0)\\ 
    &-\frac{1}{\lambda'}   P\mathcal{L}_1^\epsilon(t)\mathcal{Q}_{\lambda'}\mathcal{L}_1^0(t)P\rho(0)\nonumber\Big) .
\end{align} Here we have used also that $\mathcal{L}_1=\mathcal{L}_1^0+\mathcal{L}_1^\epsilon$ and neglected the term proportional to $\epsilon^2$.  Substituting for the definitions of the $\mathcal{L}_1$ terms, we have 

\begin{align}
     P\dot{\rho}(t)=&\sum_{\lambda'}\frac{1}{\hbar^2}\Big(\frac{1}{\lambda'} P[H_{int}^0(t),\bullet]\mathcal{Q}_{\lambda'}[H_{int}^0(t),\bullet]P\rho(0)\nonumber \\
     &+\frac{1}{\lambda'} P[H_{int}^0(t),\bullet]\mathcal{Q}_{\lambda'}[H_{int}^\epsilon(t),\bullet]P\rho(0)\\
&+\frac{1}{\lambda'}  P\mathcal[H_{int}^\epsilon(t),\bullet]\mathcal{Q}_{\lambda'}\mathcal[H_{int}^0(t),\bullet]P\rho(0)\Big).\nonumber
\end{align} Now, we trace over all of the cavity degrees of freedom as we are interested only in the mechanical degrees of freedom. Defining $\mu(t) = Tr_c[P\rho(t)]$ we have 
\begin{align}\label{eq:ProjectedLaserCoolingH}
     \mu(t)=\sum_{\lambda'}\frac{1}{\hbar^2}\Big(&Tr_c[\frac{1}{\lambda'} P[H_{int}^0(t),\bullet]\mathcal{Q}_{\lambda'}[H_{int}^0(t),\bullet]P\rho(0)]\\
    &+Tr_c[\frac{1}{\lambda'}P[H_{int}^0(t),\bullet]\mathcal{Q}_{\lambda'}[H_{int}^\epsilon(t),\bullet]P\rho(0)]\nonumber\\
    &+Tr_c[\frac{1}{\lambda'}P\mathcal[H_{int}^\epsilon(t),\bullet]\mathcal{Q}_{\lambda'}\mathcal[H_{int}^0(t),\bullet]P\rho(0)]\Big).\nonumber
\end{align} The first term yields the usual master equation for the non-driven case and the other two terms yield correction terms proportional to $\epsilon$. We may calculate term by term, using the notation

\begin{align}
F_a =&(\alpha_0^*a +\alpha_0 a^\dagger),\\
F_\Gamma =&(\Gamma+\Gamma^\dagger),\\
F_\Gamma^+ =& \frac{\epsilon}{8(n+1)}(e^{2i\omega t}\Gamma+e^{-2i\omega t}\Gamma^\dagger),\\
F_\Gamma^- =& \frac{\epsilon}{8(n-1)}(e^{-2i\omega t}\Gamma+e^{2i\omega t}\Gamma^\dagger),\\
F_\Gamma^i =& F_\Gamma^+-F_\Gamma^-,\\
H_{int}^0 =& \chi_0F_aF_\Gamma,\\
H_{int}^\epsilon =& \chi_0F_aF_\Gamma^i.
\end{align} With this, in the case of the first term of equation \eqref{eq:ProjectedLaserCoolingH}

\begin{align}
&\chi_0^2 \sum_{\lambda'}\frac{1}{\hbar^2}Tr_c[P[F_aF_\Gamma,\mathcal{Q}_{\lambda'}[F_aF_\Gamma,\mu\rho_{st}]]]\\ =&\chi_0^2 \sum_{\lambda'}\frac{1}{\hbar^2}\Big(\frac{1}{\lambda'}(Tr_c[PF_a F_\Gamma \mathcal{Q}_{\lambda'}(F_aF_\Gamma\mu\rho_{st})]\nonumber\\
&- Tr_c[P\mathcal{Q}_{\lambda'}(F_a F_\Gamma\mu\rho_{st})F_a F_\Gamma]\nonumber\\
&- Tr_c[PF_a F_\Gamma \mathcal{Q}_{\lambda'}(\mu\rho_{st}F_a F_\Gamma )]\nonumber\\
&+ Tr_c[P\mathcal{Q}_{\lambda'}(\mu\rho_{st}F_a F_\Gamma )F_a F_\Gamma])\Big).\nonumber
\end{align} Where we have assumed that $\rho(0)=\rho_{mec}\otimes\rho_{st}$, that is that the initial condition is separable. We separate the proyection operators into the mechanical and cavity parts, indicated by the appropriate sub-index

\begin{align}
    P=&P_aP_\Gamma, \\
    \mathcal{Q}_{\lambda'}=&\mathcal{Q}_{\lambda'_a}\mathcal{Q}_{\lambda'_\Gamma}.
\end{align} The mechanical parts can then be taken out of the trace and we have

\begin{align}
\chi_0^2 \sum_{\lambda'}\frac{1}{\hbar^2}&\frac{1}{\lambda'}Tr_c[P[F_aF_\Gamma,\mathcal{Q}_{\lambda'}[F_aF_\Gamma,\mu\rho_{st}]]] \\
=\chi_0^2 \sum_{\lambda'}\frac{1}{\hbar^2}&\Big(\frac{1}{\lambda'}(Tr_c[P_a \mathcal{Q}_{\lambda'_a}(F_a \rho_{st})F_a ]P_\Gamma F_\Gamma \mathcal{Q}_{\lambda'_\Gamma} (F_\Gamma\mu)\nonumber \\
&- Tr_c[P_a \mathcal{Q}_{\lambda'_a}(F_a \rho_{st})F_a ]P_\Gamma \mathcal{Q}_{\lambda'_\Gamma}( F_\Gamma\mu)F_\Gamma\nonumber\\
&- Tr_c[P_a F_a  \mathcal{Q}_{\lambda'_a}(\rho_{st}F_a) ]P_\Gamma F_\Gamma \mathcal{Q}_{\lambda'_\Gamma}(\mu F_\Gamma)\nonumber\\
&+ Tr_c[P_a  \mathcal{Q}_{\lambda'_a}(\rho_{st}F_a)F_a  ]P_\Gamma  \mathcal{Q}_{\lambda'_\Gamma}(\mu F_\Gamma)F_\Gamma\nonumber)\Big).\nonumber
\end{align} This can be written as

\begin{align}
\chi_0^2 &\sum_{\lambda'}\frac{1}{\hbar^2}\frac{1}{\lambda'}Tr_c[P[F_aF_\Gamma,\mathcal{Q}_{\lambda'}[F_aF_\Gamma,\mu\rho_{st}]]]\\
=&\chi_0^2 \sum_{\lambda'}\frac{1}{\hbar^2}\frac{1}{\lambda'}\Big( T_{1c}P_\Gamma[F_\Gamma,\mathcal{Q}_{\lambda'_\Gamma} F_\Gamma \mu]\nonumber\\
&\qquad\qquad\quad -T_{2c}P_\Gamma[F_\Gamma,\mathcal{Q}_{\lambda'_\Gamma} \mu F_\Gamma ]\Big),\nonumber
\end{align}  with

\begin{align}
T_{1c}=& Tr_c[F_a\mathcal{Q}_{\lambda'_a}F_a\rho_{st}]\nonumber\\
=&\hbar^2\abs{\alpha_0}^2 \delta_{j,1}\delta_{n,0},\\
T_{2c}=& Tr_c[F_a\mathcal{Q}_{\lambda'_a}\rho_{st}F_a]\nonumber\\
=&\hbar^2 \abs{\alpha_0}^2 \delta_{j,-1}\delta_{n,0},
\end{align} and

\begin{align}\label{CoolingBeforecoefficients}
P_\Gamma[F_\Gamma,\mathcal{Q}_{\lambda'_\Gamma} F_\Gamma \mu] =& ((\Gamma\Gamma^\dagger \mu - \Gamma^\dagger \mu  \Gamma)\delta_{l,-1} \\
&+ (\Gamma^\dagger\Gamma \mu - \Gamma \mu  \Gamma^\dagger)\delta_{l,1}),\nonumber \\
P_\Gamma[F_\Gamma,\mathcal{Q}_{\lambda'_\Gamma}  \mu F_\Gamma] =& ((\Gamma\mu\Gamma^\dagger  - \mu\Gamma^\dagger   \Gamma)\delta_{l,-1}\\ 
&+ (\Gamma^\dagger \mu\Gamma -  \mu \Gamma \Gamma^\dagger)\delta_{l,1}). \nonumber
\end{align} The other two terms in equation \eqref{eq:ProjectedLaserCoolingH} are handled in the exact same manner. Both terms yield the exact same result which then acquires a factor of 2 and we have 

\begin{align}
&Tr_c\frac{1}{\hbar^2}\Big[\frac{1}{\lambda'}P[H_{int}^0(t),\bullet]\mathcal{Q}_{\lambda'}[H_{int}^\epsilon(t),\bullet]P\rho(0)\Big]\nonumber\\
=&\chi_0^2\sum_{\lambda'}\frac{1}{\hbar^2}\Big(  \frac{2}{\lambda'}(T_{1c}P_\Gamma[F_\Gamma,\mathcal{Q}_{\lambda'_\Gamma}( F_\Gamma^{i} \mu)]\\
&\qquad \qquad \qquad -T_{2c}P_\Gamma[F_\Gamma,\mathcal{Q}_{\lambda'_\Gamma}( \mu F_\Gamma^{i} )]\Big),\nonumber
\end{align} with

\begin{align}
P_\Gamma[F_\Gamma,\mathcal{Q}_{\lambda'_\Gamma}( F_\Gamma^i \mu)] =& \epsilon(T(-\omega)\Gamma \Gamma^\dagger \mu \delta_{l,-1}-T(-\omega)\Gamma^\dagger \mu \Gamma \delta_{l,-1}\\
&+T(\omega) \Gamma^\dagger\Gamma \mu  \delta_{l,1}-T(\omega)\Gamma \mu \Gamma^\dagger \delta_{l,1})\nonumber ,
\end{align} and

\begin{align}
P_\Gamma[F_\Gamma,\mathcal{Q}_{\lambda'_\Gamma}(\mu F_\Gamma^i)]=&\epsilon(T(-\omega)\Gamma \mu \Gamma^\dagger \delta_{l,-1}-T(-\omega)\mu \Gamma^\dagger \Gamma \delta_{l,-1}\\
&+T(\omega) \Gamma^\dagger \mu \Gamma \delta_{l,1}-T(\omega)\mu \Gamma \Gamma^\dagger \delta_{l,1})\nonumber,
\end{align} where we have defined

\begin{equation}
T(\omega) = \frac{1}{8(n+1)}e^{2i\omega t}- \frac{1}{8(n-1)}e^{-2i\omega t}.
\end{equation} We approximate this function, neglecting terms of order $\frac{\epsilon}{n^2}$ as

\begin{equation}
    T(\omega)=\frac{i\sin(2\omega t)}{4n}.
\end{equation} The Kronecker delta functions apply to the eigenvalues $\lambda'$ which are, adding together the values for the cavity and the mechanical resonator

\begin{equation}
    \lambda' = i(\delta j + \nu_0 l) - \kappa(n+\frac{\abs{l}}{2}),
\end{equation} and equation \eqref{eq:ProjectedLaserCoolingH} can be re-arranged as

\begin{equation}\label{eq:MirrorCoolingEquation}
\dot{\mu}(t) = A_-D[\Gamma]\mu + A_+D[\Gamma^\dagger]\mu,
\end{equation} if we neglect a small term proportional to $\Gamma^\dagger \Gamma$, with

\begin{align}
    A_{\pm} =& A^0_{\pm} + \epsilon \sin(2\omega t) A^\epsilon_{\pm},\\
    A^0_\pm=&\frac{\chi_0^2 \abs{\alpha_0}^2}{2}\frac{\kappa}{(\delta\mp \nu_0)^2+\frac{\kappa^2}{4}}\\
    A_{\pm}^\epsilon=& \frac{\chi_0^2 \abs{\alpha_0}^2}{2}\frac{(\delta\mp\nu_0) }{n\left( \frac{{\kappa}^{2}}{4}+{{(\delta\mp\nu_0)}^{2}}\right)}\, ,
\end{align}

In the adiabatic approximation, as presented here, the mechanical
  dissipation can be incorporated into the master equation later, as it occurs on a much longer time-scale than other processes
  ($\gamma \ll \kappa, \chi_0 \abs{\alpha_0}$). By adding
  (\ref{eq:mechanical_dissipation}) to
  (\ref{eq:MirrorCoolingEquation}) we obtain

\begin{equation}
\dot{\mu} = (A_-(t)+\frac{\gamma}{2}(n_m+1))D[\Gamma]\mu + (A_+(t)+\frac{\gamma}{2}n_m)D[\Gamma^\dagger]\mu, 
\end{equation} which is the desired result.

\section{Calculation of the Covariance Matrix}\label{app:Covariance}

We follow \cite{ZollerCovMat,BarberisLC} to calculate the covariance
matrix. It is useful to first change to dimensionless position and
momentum operators

\begin{align}\label{eq:DimensionlessXP}
    X =& \sqrt{\frac{n\omega}{2}}(g(t)^* \Gamma + g(t) \Gamma^\dagger),\\
    P =& \frac{1}{\sqrt{2n\omega}}(h(t)^* \Gamma + h(t) \Gamma^\dagger),
\end{align} and define the vector

\begin{equation}
    \overline{R} = [X,P]^T.
\end{equation} The expectation value of the covariance matrix is then expressed as

\begin{equation}\label{eq:CovarianceMatrix}
    \overline{\gamma}_{i,j} = \frac{1}{2}\expval{\overline{R}_i\overline{R}_j+\overline{R}_j\overline{R}_i}-\expval{\overline{R}_i}\expval{\overline{R}_j}.
\end{equation}If the master equation \eqref{eq:LCMasterEq} can be expressed in the form

\begin{equation}\label{eq:LinearizedMasterEq}
    \dv{\mu}{t} = \sum_k \gamma_k D[\overline{L}_k\overline{R}]\mu.
\end{equation} for a pair of vectors $L_1$ and $L_2$,  a differential equation for the matrix $\overline{\gamma}$ can then be found. The differential equation for the covariance matrix is

\begin{equation}\label{eq:CovarianceDifferentialEq}
    \dv{\overline{\gamma}}{t}= \overline{H}_{eff}\overline{\gamma}+\overline{\gamma}\overline{H}^T_{eff}+\overline{J},
\end{equation} with

\begin{align}\label{eq:CovMatrixParts}
    \overline{\sigma}_{i,j}=&\frac{1}{i}[\overline{R}_i,\overline{R}_j],\\
    \overline{G}_{i,j}=& \sum_k \gamma_k (\overline{L}_k^*)_i(\overline{L}_k)_j,\\
    \overline{H}_{eff}=& 2\overline{\sigma}(Im(\overline{G})),\\
    \overline{J}=& 2\overline{\sigma}(Re(\overline{G}))\overline{\sigma}^T.
\end{align} The equation can be integrated as

\begin{align}\label{eq:CovarianceIntegralEq}
    \overline{\gamma}(t)=& e^{\int_0^tdt'\overline{H}_{eff}(t')}\overline{\gamma}(0)e^{\int_0^tdt'\overline{H}^T_{eff}(t')}\\
    &+\int_0^td\tau e^{\int_0^{t-\tau}dt'\overline{H}_{eff}(t')}\overline{J}(\tau)e^{\int_0^{t-\tau}dt'\overline{H}^T_{eff}(t')}.\nonumber
\end{align}

In order to express the master equation \eqref{eq:LCMasterEq}  in the form \eqref{eq:LinearizedMasterEq}, we require two vectors $L$ such that

\begin{align}
    \overline{L}_1\overline{R} =& \Gamma,\\
    \overline{L}_2\overline{R} =& \Gamma^\dagger.\\
\end{align} This is simple given the form of the $\Gamma$ operators

\begin{align}
    \Gamma=& \sqrt{\frac{2}{n\omega}}\frac{h(t)}{2i}X-\sqrt{2n\omega}\frac{g(t)}{2i}P,\\
    \Gamma^\dagger=& -\sqrt{\frac{2}{n\omega}}\frac{h^*(t)}{2i}X+\sqrt{2n\omega}\frac{g(t)^*}{2i}P.
\end{align}  We can then easily write

\begin{align}
     \overline{L}_1 =& \frac{1}{2i}(\sqrt{\frac{2}{n\omega}}h(t),-\sqrt{2n\omega}g(t)),\\
     \overline{L}_2 =& \frac{-1}{2i}(\sqrt{\frac{2}{n\omega}}h^*(t),-\sqrt{2n\omega}g^*(t)).
\end{align} Given the form of equation \eqref{eq:LinearizedMasterEq} we can see that

\begin{align}
    \gamma_1 =& A_-(t),\\
    \gamma_2=& A_+(t).
\end{align} We can then calculate all of the matrices in \eqref{eq:CovMatrixParts}. The commutator matrix is

\begin{equation}
    \overline{\sigma}_{i,j}=\mqty(0&1\\-1&0),
\end{equation} and so

\begin{align}
    \overline{G}_{1,1} =& A_-(\overline{L}^*_1)_1(\overline{L}_1)_1+A_+(\overline{L}^*_2)_1(\overline{L}_2)_1,\\
    \overline{G}_{1,2} =& A_-(\overline{L}^*_1)_1(\overline{L}_1)_2+A_+(\overline{L}^*_2)_1(\overline{L}_2)_2,\\
    \overline{G}_{2,1} =& A_-(\overline{L}^*_1)_2(\overline{L}_1)_1+A_+(\overline{L}^*_2)_2(\overline{L}_2)_1,\\
    \overline{G}_{2,2} =& A_-(\overline{L}^*_1)_2(\overline{L}_1)_2+A_+(\overline{L}^*_2)_2(\overline{L}_2)_2.
\end{align} This allows us to write expressions for $\overline{H}_{eff}$ and $\overline{J}$

\begin{align}
    \overline{H}_{eff}=& (A_+-A_-)\overline{I},\\
    \overline{J}=&(A_++A_-)\overline{I},
\end{align} with 

\begin{equation}
    \overline{I}=\mqty(1&0\\0&1)\, .
  \end{equation}
  After separating the coefficients $A_\pm$ as in equation
  \eqref{eq:ACoefficients}, we begin the integration process. We
  begin with the integrals appearing in the exponentials in equation
  \eqref{eq:CovarianceIntegralEq}

\begin{align}
    \int_0^tdt'\overline{H}_{eff}(t') =&(A^0_+-A^0_-)t \overline{I}\nonumber\\
    &+\frac{\epsilon(1- \cos(2\omega t))}{2\omega}(A^\epsilon_+-A^\epsilon_-)\overline{I},\\
    \int_0^{t-\tau}dt'\overline{H}_{eff}(t') =& (A^0_+-A^0_-)(t-\tau)\overline{I}\nonumber\\
    &+\frac{\epsilon(1- \cos(2\omega (t-\tau)))}{2\omega}(A^\epsilon_+-A^\epsilon_-)\overline{I}
\end{align} Remembering that $\overline{H}_{eff}=\overline{H}^T_{eff}$ due to symmetry. If we define

\begin{align}
    \overline{A}_0 =&(A^0_+-A^0_-)\overline{I},\\
    \overline{A}_\epsilon =&\frac{\epsilon }{2\omega}(A^\epsilon_+-A^\epsilon_-)\overline{I}.
\end{align} We may then write \eqref{eq:CovarianceIntegralEq} as
\begin{align}
    \overline{\gamma(t)} =& e^{\overline{A}_0t+\overline{A}_\epsilon (1-\cos(2\omega t))}\overline{\gamma}(0) e^{\overline{A}_0t+\overline{A}_\epsilon (1-\cos(2\omega t))}\\
    &+\int_0^td\tau e^{\overline{A}_0(t-\tau)+\overline{A}_\epsilon(1- \cos(2\omega (t-\tau)))}\nonumber\\
    &\qquad\overline{J}(\tau) e^{\overline{A}_0(t-\tau)+\overline{A}_\epsilon(1- \cos(2\omega (t-\tau)))}.\nonumber
\end{align} Since both $\overline{A}_0$ and $\overline{A}_\epsilon$ are proportional to $\overline{I}$, they commute with any $2\times2$ matrix and equation \eqref{eq:CovarianceIntegralEq} simplifies to

\begin{align}
    &\overline{\gamma(t)} = e^{2(\overline{A}_0t+\overline{A}_\epsilon (1-\cos(2\omega t)))}\overline{\gamma}(0)\\ 
    &+\int_0^td\tau e^{2(\overline{A}_0(t-\tau)+\overline{A}_\epsilon (1-\cos(2\omega (t-\tau))))}\overline{J}(\tau).\nonumber
\end{align} We then tackle the remaining integrals. We may separate the matrix $\overline{J}$ as

\begin{align}
    \overline{J}=&\overline{J}_0+\sin(2\omega t)\overline{J}_\epsilon\\
    \overline{J}_0 =&(A_-^0+A_+^0)\overline{I},\\
     \overline{J}_\epsilon =&\epsilon(A_-^\epsilon+A_+^\epsilon)\overline{I}.
\end{align} We also employ the approximation 

\begin{equation}
    e^{2(\overline{A}_\epsilon(1- \cos(2\omega t)))} \approx \overline{I} + 2\overline{A}_\epsilon(1- \cos(2\omega t)).
\end{equation} So that, up to first order in $\epsilon$

\begin{align}
    \int_0^td\tau &e^{2(\overline{A}_0(t-\tau)+\overline{A}_\epsilon (1-\cos(2\omega (t-\tau))))}\overline{J}(\tau)\nonumber\\
    =& \int_0^td\tau e^{2\overline{A}_0(t-\tau)}\overline{J}_0\\
    &+\int_0^td\tau e^{2\overline{A}_0(t-\tau)}\sin(2\omega \tau)\overline{J}_\epsilon \nonumber\\
   &-2\int_0^td\tau e^{2\overline{A}_0(t-\tau)} \overline{A}_\epsilon \cos(2\omega (t-\tau))\overline{J}_0 \nonumber\\
   &+2\int_0^t d\tau e^{2\overline{A}_0(t-\tau)}\overline{A}_\epsilon \overline{J}_0.\nonumber
\end{align} The integral that is not proportional to $\epsilon$ yields

\begin{align}
 \int_0^td\tau e^{2\overline{A}_0(t-\tau)}\overline{J}_0 =& \frac{1}{2}\frac{A_+^0+A_-^0}{A_+^0-A_-^0}e^{2(A_+^0-A_-^0)t}\overline{I}\nonumber\\ 
 &+\frac{1}{2}\frac{A_+^0+A_-^0}{A_-^0-A_+^0}\overline{I}.
 \end{align} And the terms proportional to $\epsilon$ are
 
 \begin{align}
 &\int_0^td\tau e^{2\overline{A}_0(t-\tau)}\sin(2\omega \tau)\overline{J}_\epsilon\\
 =&\frac{1}{2}\epsilon(A_+^\epsilon+A_-^\epsilon)\frac{\omega e^{2(A_+^0-A_-^0)t}}{(A_+^0-A_-^0)^2+\omega^2}\overline{I}\nonumber\\
 &-\frac{1}{2}\epsilon(A_+^\epsilon+A_-^\epsilon)\frac{(A_+^0-A_-^0)\sin(2\omega t)+\omega\cos(2\omega t)}{(A_+^0-A_-^0)^2+\omega^2}\overline{I}\nonumber.
\end{align} And

\begin{align}
  -&\int_0^td\tau e^{2\overline{A}_0(t-\tau)} 2\overline{A}_\epsilon \cos(2\omega (t-\tau))\overline{J}_0=\nonumber\\
  &\frac{\epsilon}{2\omega}\frac{(A_+^\epsilon-A_-^\epsilon)(A_+^0-A_-^0)(A_+^0+A_-^0)}{(A_+^0-A_-^0)^2+\omega^2}\overline{I}\\
  &-\frac{\epsilon}{2\omega}(A_+^\epsilon-A_-^\epsilon)(A_+^0+A_-^0)\frac{e^{2(A_+^0-A_-^0)t}}{(A_+^0-A_-^0)^2+\omega^2}\Big(\nonumber\\
  &\quad (A_+^0-A_-^0)\cos(2\omega t) + \omega\sin(2\omega t)\Big)\overline{I}\nonumber.
\end{align} And

\begin{align}
 2&\int_0^t d\tau e^{2\overline{A}_0(t-\tau)}\overline{A}_\epsilon \overline{J}_0 =  2\overline{A}_\epsilon \overline{J}_0(\frac{e^{2\overline{A}_0 t}}{2\overline{A}_0}-\frac{1}{2\overline{A}_0})\\=&\frac{\epsilon}{2\omega}(\frac{e^{2(A_+^0-A_-^0)t}(A_+^\epsilon - A_-^\epsilon)(A_-^0+A_+^0)}{A_+^0-A_-^0})\overline{I}\nonumber\\
 &-\frac{\epsilon}{2\omega}(\frac{(A_+^\epsilon - A_-^\epsilon)(A_-^0+A_+^0)}{A_+^0-A_-^0})\overline{I}\nonumber .
\end{align}
After a long enough time, if the parameters are chosen to favor cooling, where $A_-> A_+$, all of the exponential terms proportional to $e^{2(A_+^0-A_-^0)t}$, including the initial condition,  drop out and equation \eqref{eq:CovarianceIntegralEq} simplifies to

\begin{align}
    \overline{\gamma(t)} =& \frac{1}{2}\frac{A_+^0+A_-^0}{A_-^0-A_+^0}\overline{I}\\
    &-\frac{1}{2}\epsilon(A_+^\epsilon+A_-^\epsilon)\frac{(A_+^0-A_-^0)\sin(2\omega t)+\omega\cos(2\omega t)}{(A_+^0-A_-^0)^2+\omega^2}\overline{I}\nonumber\\
    &+\frac{\epsilon}{2\omega}\frac{(A_+^\epsilon-A_-^\epsilon)(A_+^0-A_-^0)(A_+^0+A_-^0)}{(A_+^0-A_-^0)^2+\omega^2}\overline{I}\nonumber\\
    &-\frac{\epsilon}{2\omega}\frac{(A_+^\epsilon - A_-^\epsilon)(A_-^0+A_+^0)}{(A_+^0-A_-^0)}\overline{I}\nonumber .
\end{align} To obtain $\expval{m}$ we must then simply take the trace

\begin{align}
    Tr[\overline{\gamma(t)}] =&\frac{A_+^0+A_-^0}{A_-^0-A_+^0}\\
    &-\epsilon(A_+^\epsilon+A_-^\epsilon)\frac{(A_+^0-A_-^0)\sin(2\omega t)}{(A_+^0-A_-^0)^2+\omega^2}\nonumber\\
    &-\epsilon(A_+^\epsilon+A_-^\epsilon)\frac{\omega\cos(2\omega t)}{(A_+^0-A_-^0)^2+\omega^2}\nonumber\\
    &+\frac{\epsilon}{\omega}\frac{(A_+^\epsilon-A_-^\epsilon)(A_+^0-A_-^0)(A_+^0+A_-^0)}{(A_+^0-A_-^0)^2+\omega^2}\nonumber\\
    &-\frac{\epsilon}{\omega}\frac{(A_+^\epsilon - A_-^\epsilon)(A_-^0+A_+^0)}{(A_+^0-A_-^0)}\nonumber .
\end{align} This is the desired result.

\bibliographystyle{unsrt}
\bibliography{Bib}

\begin{thebibliography}{10}

\bibitem{BraginskiiOG}
V.~B. Braginskii{,} A.~B. Manukin{,} and M.~Yu. Tikhonov.
\newblock Investigation of dissipative poderomotive effects of electromagnetic
  radiation.
\newblock {\em Soviet Physics JETP}, 31(5):829, November 1970.

\bibitem{WeissOIT}
Stefan Weis, R{\'e}mi Rivi{\`e}re, Samuel Del{\'e}glise, Emanuel Gavartin,
  Olivier Arcizet, Albert Schliesser, and Tobias~J. Kippenberg.
\newblock Optomechanically induced transparency.
\newblock {\em Science}, 330(6010):1520--1523, 2010.

\bibitem{VogelOT}
M.~Vogel, C.~Mooser, K.~Karrai, and R.~J. Warburton.
\newblock Optically tunable mechanics of microlevers.
\newblock {\em Applied Physics Letters}, 83(7):1337--1339, 2003.

\bibitem{CohadonCM}
P.~F. Cohadon, A.~Heidmann, and M.~Pinard.
\newblock Cooling of a mirror by radiation pressure.
\newblock {\em Phys. Rev. Lett.}, 83:3174--3177, Oct 1999.

\bibitem{CorbittOC}
Thomas Corbitt, Christopher Wipf, Timothy Bodiya, David Ottaway, Daniel Sigg,
  Nicolas Smith, Stanley Whitcomb, and Nergis Mavalvala.
\newblock Optical dilution and feedback cooling of a gram-scale oscillator to
  6.9 mk.
\newblock {\em Phys. Rev. Lett.}, 99:160801, Oct 2007.

\bibitem{SchliesserRPC}
A.~Schliesser, P.~Del'Haye, N.~Nooshi, K.~J. Vahala, and T.~J. Kippenberg.
\newblock Radiation pressure cooling of a micromechanical oscillator using
  dynamical backaction.
\newblock {\em Phys. Rev. Lett.}, 97:243905, Dec 2006.

\bibitem{LCNooshi}
I~Wilson-Rae, N~Nooshi, J~Dobrindt, T~J Kippenberg, and W~Zwerger.
\newblock Cavity-assisted backaction cooling of mechanical resonators.
\newblock {\em New Journal of Physics}, 10(9):095007, sep 2008.

\bibitem{ManciniOC}
Stefano Mancini, David Vitali, and Paolo Tombesi.
\newblock Optomechanical cooling of a macroscopic oscillator by homodyne
  feedback.
\newblock {\em Physical Review Letters}, 80, 02 1998.

\bibitem{MarquardtQTOQ}
Florian Marquardt{,} A. A. Clerk{,}S.~M. Girvin.
\newblock Quantum theory of optomechanical cooling.
\newblock {\em Journal of Modern Optics}, 55(19-20):3329--3338, 2008.

\bibitem{KippenberCO}
Markus Aspelmeyer, Tobias~J. Kippenberg, and Florian Marquardt.
\newblock Cavity optomechanics.
\newblock {\em Rev. Mod. Phys.}, 86:1391--1452, Dec 2014.

\bibitem{JockelMR}
Stephan Camerer, Maria Korppi, Andreas J\"ockel, David Hunger, Theodor~W.
  H\"ansch, and Philipp Treutlein.
\newblock Realization of an optomechanical interface between ultracold atoms
  and a membrane.
\newblock {\em Phys. Rev. Lett.}, 107:223001, Nov 2011.

\bibitem{AranasSlowModulation2016}
P.~F.~Barker E.~B.~Aranas, P. Z. G.~Fonseca and T.~S. Monteiro.
\newblock Split-sideband spectroscopy in slowly modulatedoptomechanics.
\newblock {\em New J. Phys}, 18:113021, 2016.

\bibitem{YinNonlinearEffectsModulated2017}
Tai-Shuang Yin, Xin-You L\"u, Li-Li Zheng, Mei Wang, Sha Li, and Ying Wu.
\newblock Nonlinear effects in modulated quantum optomechanics.
\newblock {\em Phys. Rev. A}, 95:053861, May 2017.

\bibitem{ZhangModulationSqueezing2018}
Zhu-Cheng Zhang, Yi-Ping Wang, Ya-Fei Yu, and Zhi-Ming Zhang.
\newblock Quantum squeezing in a modulated optomechanical system.
\newblock {\em Opt. Express}, 26(9):11915--11927, Apr 2018.

\bibitem{MalzBiChromaticallyDriven2016}
Daniel Malz and Andreas Nunnenkamp.
\newblock Floquet approach to bichromatically driven cavity-optomechanical
  systems.
\newblock {\em Phys. Rev. A}, 94:023803, Aug 2016.

\bibitem{MariGentleModulation2009}
A.~Mari and J.~Eisert.
\newblock Gently modulating optomechanical systems.
\newblock {\em Phys. Rev. Lett.}, 103:213603, Nov 2009.

\bibitem{BarberisLC}
Marc Bienert and Pablo Barberis-Blostein.
\newblock Optomechanical laser cooling with mechanical modulations.
\newblock {\em Phys. Rev. A}, 91:023818, Feb 2015.

\bibitem{HanngiFM}
Sigmund Kohler, Thomas Dittrich, and Peter Hanggi.
\newblock Floquet-markov description of the parametrically driven, dissipative
  harmonic quantum oscillator.
\newblock {\em Physical Review E}, 55, 09 1998.

\bibitem{BrownPT}
Lowell~S. Brown.
\newblock Quantum motion in a paul trap.
\newblock {\em Phys. Rev. Lett.}, 66:527--529, Feb 1991.

\bibitem{WardFT}
M.J. Ward.
\newblock {\em Industrial Mathematics, Lecture Notes, Dept. of Mathematics}.
\newblock Unpublished, Univ. of British Columbia, Vancouver, B.C., Canada,
  2008.

\bibitem{PiatekME}
Marcin Piatek and Artur~R. Pietrykowski.
\newblock Classical irregular blocks, hill's equation and pt-symmetric periodic
  complex potentials.
\newblock {\em Journal of High Energy Physics}, 2016(7):131, Jul 2016.

\bibitem{ZollerCovMat}
M.~Wallquist, K.~Hammerer, P.~Zoller, C.~Genes, M.~Ludwig, F.~Marquardt,
  P.~Treutlein, J.~Ye, and H.~J. Kimble.
\newblock Single-atom cavity qed and optomicromechanics.
\newblock {\em Phys. Rev. A}, 81:023816, Feb 2010.

\bibitem{ParkSidebandCryogenic2009}
Young-Shin Park and Hailin Wang.
\newblock Resolved-sideband and cryogenic cooling of an optomechanical
  resonator.
\newblock {\em Nature Physics}, 5(7):489--493, Jul 2009.

\bibitem{PetersonMicromechanicalMembraneBackactionLimit2016}
R.~W. Peterson, T.~P. Purdy, N.~S. Kampel, R.~W. Andrews, P.-L. Yu, K.~W.
  Lehnert, and C.~A. Regal.
\newblock Laser cooling of a micromechanical membrane to the quantum backaction
  limit.
\newblock {\em Phys. Rev. Lett.}, 116:063601, Feb 2016.

\bibitem{EnglertDB}
Hans-J\"urgen Briegel and Berthold-Georg Englert.
\newblock Quantum optical master equations: The use of damping bases.
\newblock {\em Phys. Rev. A}, 47:3311--3329, Apr 1993.

\bibitem{CarmichaelQO}
Howard Carmichael.
\newblock {\em Statistical Methods in Quantum Optics, Vol 1}, volume~1 of {\em
  Theoretical and Mathematical Physics}.
\newblock Springer, 1 edition, 1999.

\end{thebibliography}

\end{document}